\documentstyle[sprocl]{article}

\input{psfig.sty}

\def\Journal#1#2#3#4{{#1} {\bf #2}, #3 (#4)}

\def\araa{\it A.R.A.A., }

\def\PRL{\it Phys. Rev. Lett.}

\def\be{\begin{equation}}
\def\ee{\end{equation}}
\def\bea{\begin{eqnarray}}
\def\eea{\end{eqnarray}}

\begin{document}

\title{ASTROPHYSICAL POLARIMETRY OF COSMOLOGICAL SOURCES}

\author{ROBERT ANTONUCCI}

\address{University of California, Physics Department, Santa Barbara,\\
 CA 93106-9530, USA\\E-mail: antonucci@physics.ucsb.edu}

%


\maketitle
\abstracts{Kostelecky and Mewes have recently shown that sensitive constraints
can be placed on some aspects of Lorentz symmetry violation using certain
astronomical data on high-redshift sources.  Here, I introduce that
data in its astronomical context, making it clear that these data
are robust and accurate for their purpose.  In particular, I explain
that spatially extended scattered light from obscured quasars leads
to a centrosymmetric scattering polarization, with polarization
position angle independent of wavelength.  Evidentally, these relationships
aren't spoiled by propagation effects as the photons cross the
universe.}

\section{Introduction}

Many astronomical objects are hidden from direct view by clouds of gas
accompanied by small solid particles called dust by astronomers.
If scattering
particles are also present, a halo of scattered light is observable,
with tangential position angles in a centro-symmetric pattern around the
direction on the sky of the hidden object.  For example, during a total
solar eclipse, a beautiful outer atmosphere becomes visible (the ``corona")
and much of this light is simply Thomson-scattered from the unseen bright
solar photosphere.
For Thomson scattering, spectral features
are broadened somewhat because of the electrons' thermal motions.

A {\it radio galaxy}\/ is a galaxy with powerful radio emission, typically
from two ``lobes" on $10^6$ light year scales, symmetrically placed about
the galaxy.  The detected continuum, which is spatially resolved
on the sky, is mostly starlight from the host
galaxies.   By contrast, a {\it quasar}\/ is a pointlike optical
continuum source with a
similar radio manifestation.  This powerful and variable continuum source
{\it may}\/ be thermal emission from gas heated as it falls toward the event
horizon of a supermassive black hole.

The quasar pointlike optical sources are surrounded by fuzzy light, which is
in fact mostly starlight from its host galaxy.
It is now known that the powerful distant radio galaxies and quasars
differ only in orientation with respect to the line of sight!
In particular the powerful
radio galaxies contain quasars hidden from direct view by dust clouds.
We detect
the hidden quasars via their halos of scattered light, just as in the case
of the solar corona.  The observed perfect tangential and centrosymmetric
polarization patterns are evidentally
unaffected by the long traverse through the universe.

Data on the material in this review are analyzed quantitatively by
Wardle et al.\ (1997).

\begin{figure}
\psfig{file=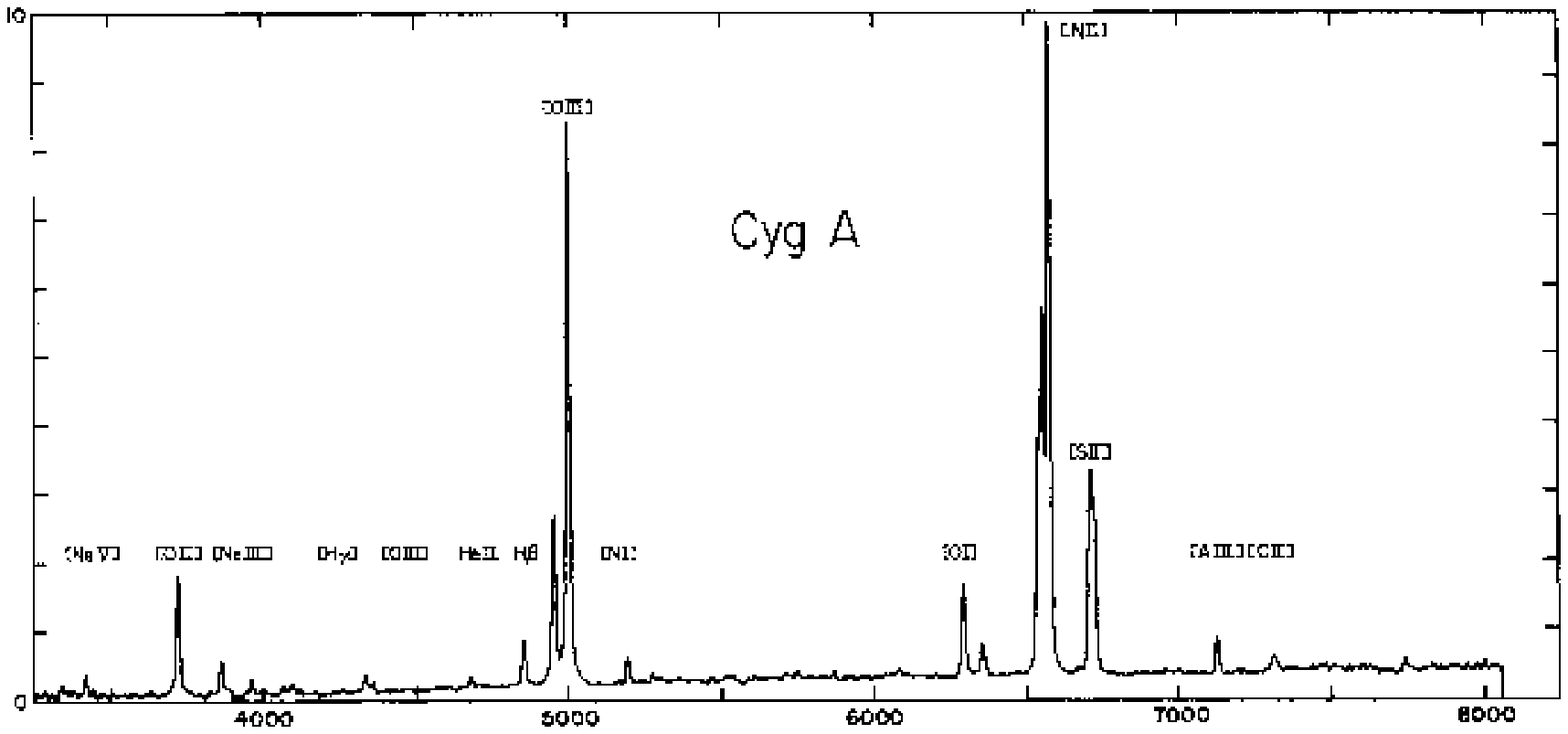,width=12cm}

\caption{Optical spectrum of the radio galaxy Cygnus A, taken from
Osterbrock 1993, Pub ASP 95, 12}

\end{figure}

\section{The Unified Model}
Let's look closely at the optical spectra of radio galaxies and quasars.
Figure~1 shows a spectrum of the typical powerful radio galaxy Cygnus A,
with a modest
redshift of 0.07:  the continuum light
is mostly starlight.  The emission lines however must come from clouds of
ionized gas.  The emission lines are due to recombination or collisional
excitation
of species photoionized by the quasar continuum.  The spectrum is quite
different from that of laboratory gases:  the transitions which violate
electric dipole selection rules are quite strong.  This indicates a very low
density gas.  In such a gas, every time an excited state is populated,
whether or not it's metastable, a photon
results from the subsequent spontaneous radiative de-excitation.
The Einstein ``A" coefficient,
which is many orders of magnitude smaller for the forbidden transitions than
for the permitted ones, makes no difference.  Every excitation results in a
``bankable" photon, whether the decay time is very long or very short.

The finite spectral width of both kinds of emission lines
comes from bulk motions of the ionized gas clouds, and indicate
internal motions within the emitting region of $\sim1,000$ km/sec.

Figure~2 shows the spectrum of the quasar (or ``broad line radio galaxy,"
for any astronomers reading this) called 3C382.
It consists of some starlight,
the same type of low-density emission lines
seen in the radio galaxy Cygnus A
above, plus two more components.  One is a mysterious continuum component,
which is powerful and variable and {\it may}\/ come from thermal radiation by
optically thick matter accreting onto a supermassive black hole;  and
the other is the broad bases {\it on the permitted lines only}\/.  These must
come from a family of ionized gas clouds with a $\sim10,000$ km/sec velocity
dispersion, and a relatively high density so that the permitted lines
are much much stronger than the forbidden lines.  At these densities
the gas is like that in the laboratory:  the excited states are well
populated for both permitted and forbidden transitions, so their ratio
is determined by the Einstein A's in this case.

\begin{figure}
\psfig{file=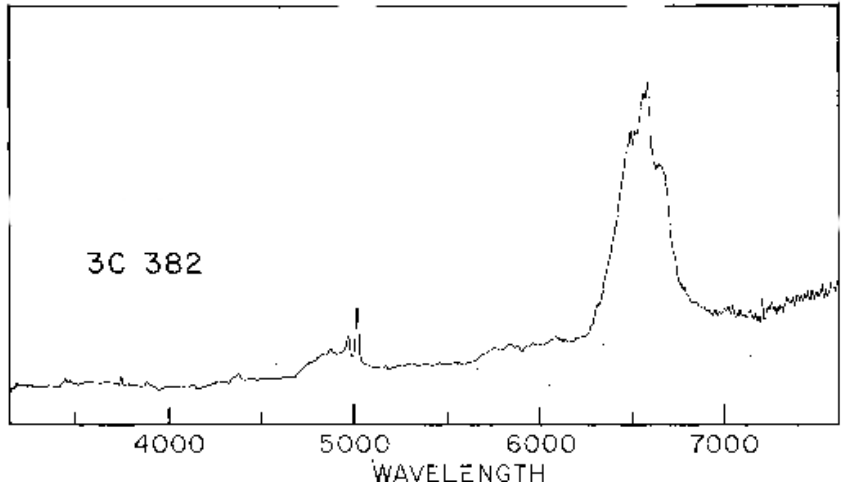,width=12cm}

\caption{Optical spectrum of the quasar (or ``broad line radio galaxy'')
3C 382, from Osterbrock, Koski and Phillips, Ap J 206, 898}

\end{figure}

During the 1980's it was deduced from polarimetry and other data that
the latter two components, the variable (and thus extremely compact)
continuum source and the broad-emission-line region, are in fact present
in powerful radio galaxies as well as in quasars, and can be
detected via scattered polarized light.
The electric vectors are always perpendicular to the radio source
symmetry axes, and thus the photons' last flights before being scattered
into the line of sight were {\it along}\/ the radio axes.  From the point of
view of the nuclear light, the other directions are blocked by a
torus-shape collection of dusty (neutral) gas clouds. Figure~3 shows a cartoon.

The scattered light around a radio galaxy manifests as a bipolar
reflection region (Fig.~4)
and a polarized light spectrum identical to the total-flux
spectra of quasars,
complete with the nuclear continuum and the broad wings on the
permitted emission lines.

The Unified Model for active galactic nuclei and quasars is reviewed
in Antonucci 1993.

\begin{figure}
\psfig{file=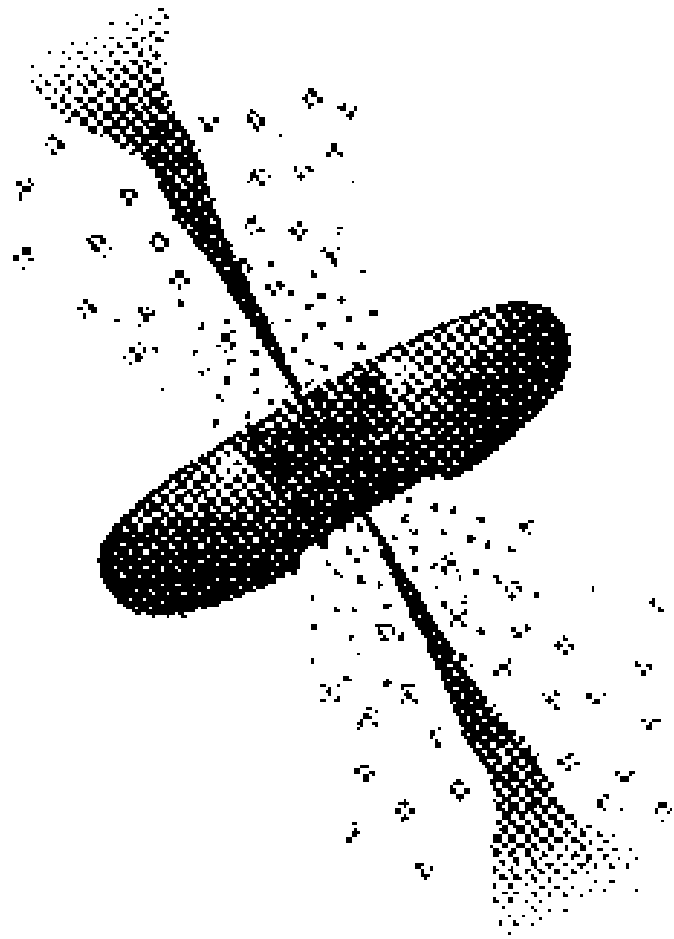,width=12cm}

\caption{Cartoon showing the obscuring torus, polar scattering regions,
and radio jets, from Urry and Padovani 1995 Pub ASP 107, 803}

\end{figure}

\begin{figure}
\psfig{file=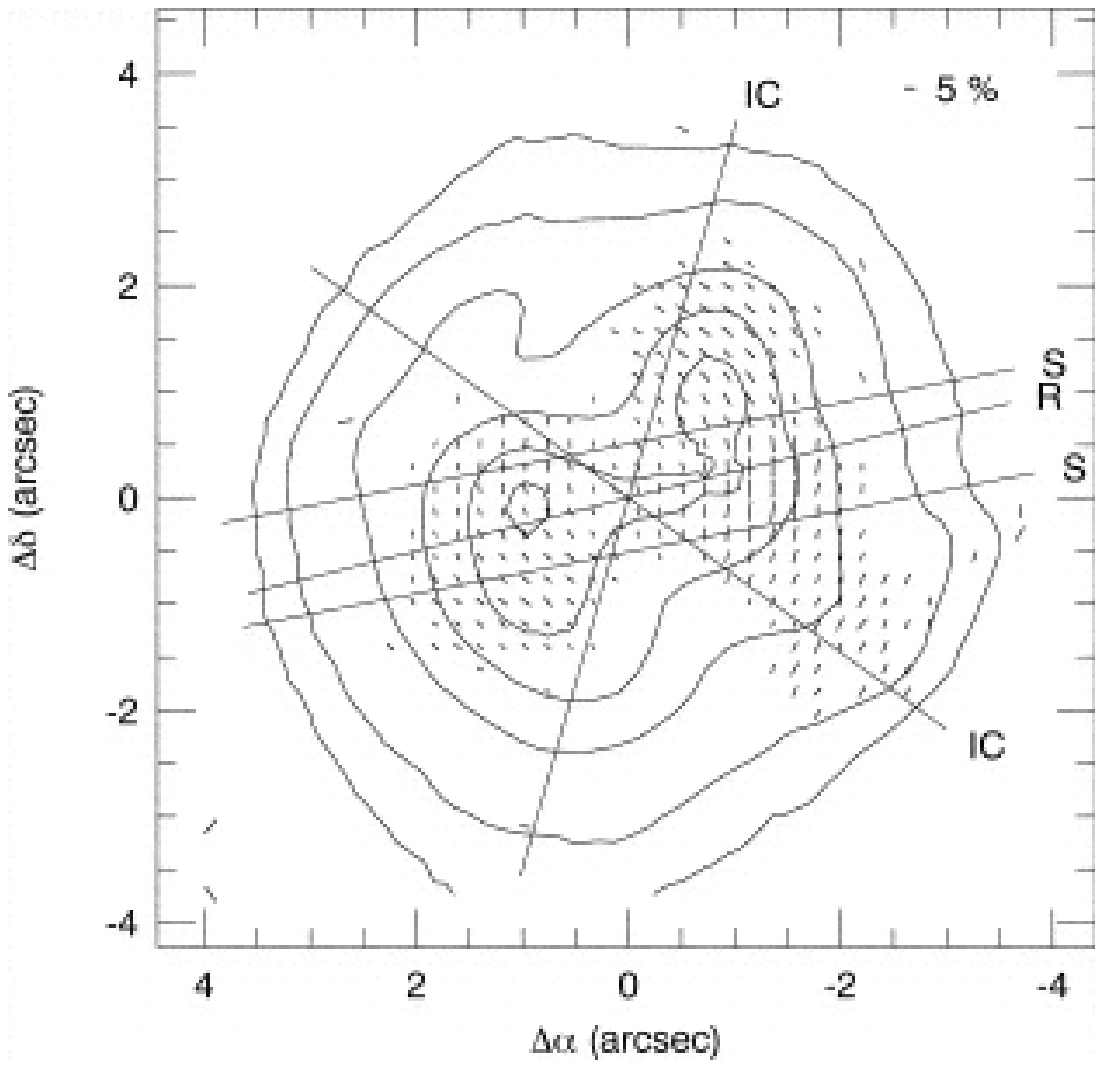,width=12cm}

\caption{Polarization image of the Cygnus A radio galaxy, showing the bipolar
reflection nebula.  From Ogle et al 1997 Ap J 482, L37}

\end{figure}

\section{Implications for Particle Physicists}

These scattering regions are visible at redshifts up
to $>2$, though the data aren't as pretty as for the nearby case
shown here.
The scatterers are sometimes small dust particles and sometimes free
electrons, or a combination of the two.  The point is, the electric
vectors are all exactly perpendicular to the direction to a single
point, where the hidden quasar lies.  Any propagation effects that
would spoil this perpendicular relationship are tightly constrained
in size.  One example is foreground gravitational lensing, which
rotates the radius vectors due to shear, but leaves the electric
vectors of the scattered light unchanged.  Another is the hypothetical
Lorentz-symmetry-violating effect discussed by Kostelecsky and Mewes.
(Their constraint depends on the ratio of the wavelength of observation
to the source distance, so will become several orders of magnitude
more powerful when X-ray polarization can be used.)

Finally let me apprise you of a related effect which can sometimes
be used for the same purpose.  {\it It is less accurate and robust.}\/
Consider the radio maps themselves.  The radio photons are also
polarized, but by a different effect:  here the photons are
intrinsically polarized because they derive from the synchrotron
process.  Since the double-lobed radio sources are axisymmetric
to zeroth order, the net radio polarization integrated over the
entire source tends to be either parallel or perpendicular to the
structural axis.   This just follows from the overall approximate
axisymmetry of the radio morphology.  Its polarization
is not trivially understood
from first principles as for the scattered optical light.  And because
it depends on the detailed ``gastrophysics" or environmental
``weather," it isn't
precise;  for some objects the polarization angle isn't related
to the symmetry axis at all.
However, statistically we can compare the radio structural axes
with the integrated
radio polarization electric vector position angles, and limit
deviations in a statistical sense:  clearly if all the polarization
angles (or all of the structural
axes) were rotated by over a radian by some propagation effect,
the statistical correlation with the radio axes
would be destroyed.

\section*{Acknowledgments}
My polarimetry work is partially funded by NSF grants NSF AST96-17160 and NSF
AST00-98719.  Also, I thank Pat Ogle for Fig.~4 and for the Wardle et
al.\ reference; Don Osterbrock for Figs.~1 and 2; and M.~Urrey and Padovani
for Fig.~3.

%

\end{document}